\documentclass[12pt,a4paper]{article}

\newcommand{\lsim}{\raisebox{0.45ex}{$<$}
                      \hspace{-0.70em}
                      \raisebox{-0.55ex}{$\sim$}}

\usepackage{graphicx}
\usepackage{amssymb}
\usepackage{amsmath}

\begin{document}

\title{Comparison of Analytical Methods of $E1$ Strength Calculations
in Middle and Heavy Nuclei}

\author{V.A. Plujko$^{1,2}$, O.O. Kavatsyuk$^{1}$ }

\date{}

\maketitle

\vspace{-1cm}
\begin{center}
$^{1}$Taras Shevchenko National  University,
Glushkova Str. 6, Kiev, Ukraine \\
$^{2}$Institute for Nuclear Research,
Prosp. Nauki 47, Kiev, Ukraine\\
E-mail: plujko@univ.kiev.ua
\end{center}
\vspace{0.1cm}

\begin{center}
\parbox{0.8\textwidth}{\small Simple analytical models for E1
strength function calculations of the $\gamma$-decay are investigated.
The MLO and GFL models (\cite{plu99a}-\cite{MD2000}) are recommended
as the best models for $E1$ gamma-decay strength function calculations.}
\end{center}

\section{Introduction}
 Gamma-emission is one of the most universal channel of the nuclear
deexcitation processes which can accompany any nuclear reaction.
Gamma-decay as well as photoabsorption can be described through
the use of the radiative strength (RS) functions \cite{Bart73,Lone86}.
The electron- positron decay is also
depended on shape of these functions. There
are radiative strength functions of two types.
The photoexcitation strength function  is connected with
cross-section of the gamma-ray absorption.
The average
radiative width $\Gamma_{\lambda}$  is determined by a gamma- decay (downward)
strength function $\overleftarrow{f}_{\lambda}$,
$ \overleftarrow{f}_{\lambda}\equiv
\Gamma_{\lambda}(\epsilon_{\gamma })
\rho (U)/3 \epsilon_{\gamma}^{3}
\rho (U-\epsilon_{\gamma})$,
where $\rho (U)$ is total density of the initial states
in heated nuclei at initial excitation energy $U$ and $\epsilon_{\gamma}$
is the gamma-ray energy.
Therefore the RS function is important constituent of the
compound nucleus model calculations of capture cross sections,
gamma-ray production spectra, isomeric state populations and of
competition between gamma-ray and particle emission. The gamma-ray
strengths are ingredient of time-consuming  calculations of
different nuclear processes, for these reasons  simple closed-form
expressions are preferred for strength functions. The approaches
based  on recent achievements in nuclear studies are also
preferable to improve the reliability of the RS function
expressions.

Dipole electric gamma-transitions are  dominant, when  they occur simultaneously
with  transitions of other multipolarities and types. According to the Brink
hypothesis \cite{Brink55,Axel62}, the Lorentzian line shape with the energy-
independent width  (SLO model) is  used widely for calculations of the dipole RS
function. This approach is probably  the most appropriate simple method for a
description of the photoabsorption data in medium-weight and heavy
nuclei \cite{Lone86,BerFultz,db}. In the case of the gamma-emission, the SLO model
strongly underestimates \cite{Pop82} the gamma-ray spectra at low energies
$\epsilon_{\gamma}~\lsim~1 MeV$. A global description
of the gamma-spectra by the Lorentzian can be obtained in the range
$1~\lsim~\epsilon_{\gamma}~\lsim~8 MeV$ only when the giant dipole resonance
(GDR) parameters are   inconsistent with that ones for
photoabsorption data. On the whole, the SLO approach overestimates the
integral experimental data like the capture cross sections and the average
radiative widths in heavy
nuclei (\cite{Lone86,Mc81}-\cite{RIPL}).

The first model with correct description of the $E1$ strengths at the energies
$\epsilon_{\gamma}$ near zero was proposed in Ref.\cite{KMF83}. Thereafter
an enhanced generalized Lorentzian model (EGLO) was developed and analyzed
in Refs.\cite{kuc93,RIPL} for a unified description of the low-energetic and
integral data. The EGLO radiative strength function consists of two
components for spherical nuclei: the Lorentzian with the energy and
temperature dependent empirical width and additional term \cite{KMF83} corresponding
to zero value of $\gamma$~- ray energy.
The EGLO method reproduces the experimental data on gamma-decay rather
well in the mass region $A= 50 \div 200$.
It should be noted that the EGLO and SLO  expressions for the
gamma-decay strength function of heated nuclei are in fact the parameterizations
of the experimental data. They are in contradiction with some aspect of the
recent theoretical studies, specifically:

{\it 1)} the shapes of the radiative strengths within  these approaches
do not consistent with  general relation between the RS function for
gamma-decay of  heated nuclei and the imaginary part of the  nuclear
response function on electromagnetic field \cite{ER1993,plu90};

{\it 2)} the damping width of the EGLO model is proportional to
collisional component of  zero sound damping width in the infinite
Fermi- liquid when  the  collisional (two-body) relaxation is taken into
account only. However the important contribution to the total width in
nuclei is also given by the fragmentation (one-body) width arising from
nucleon motion in  self-consistent mean field \cite{BB94}. This width is
almost independent of the nuclear temperature and is not included in
EGLO model. Note that the SLO-model width is independent of energy; that is,
it has properties of the fragmentation width but it is not allowing for
collisional damping.

Recently new closed-form models for the E1 strength were proposed in Refs.
(\cite{plu99a}-\cite{MD2000}) which  avoid some of these
defects at least in approximate way.

Specifically, the approach considered in
Refs.(\cite{plu99a}-\cite{plu2001})
is in line with detailed balance principle \cite{SK1986}. This method was named
previously as the thermodynamic pole approximation.
It is denoted below as  the modified Lorentzian (MLO) approach,
because a resulting expression for the dipole RS function  has a Lorentzian shape
scaled by an enhanced factor. An expression for energy-dependent damping
width of the MLO model includes also component corresponding to fragmentation
contribution in some simplified way.
Damping width appearing in the generalized Fermi liquid (GFL) model \cite{MD2000}
of the E1 strength has fragmentation component too; it is taken as
resulted from the dipole-quadrupole interaction. Below we investigate
these new approaches to compare their
with EGLO and SLO models.


\section{Simple closed-form models for radiative \\
strength function calculations}

The expression for dipole radiative
strength function has the following form in SLO model
\begin{equation}
\vspace{-1.0mm}
\label{SLO} \overleftarrow{f}_{SLO}(\epsilon_{\gamma}) = 8,674
\cdot 10^{-8} \sigma_{r} \Gamma_{r} \frac{\epsilon_{\gamma}
\Gamma_{r}}{ ( \epsilon_{\gamma}^2 - E_{r}^2)^2 +(\Gamma_{r}
\epsilon_{\gamma})^2},\ \ \ \ MeV^{-3},
\end{equation}
where $\sigma_{r}$ (in {\it mb}), $\Gamma_{r}$ and $E_{r}$
(in {\it MeV}) are GDR parameters.

The expression for gamma-decay dipole RS function
within EGLO model \cite{KMF83} is the following,
$$\overleftarrow{f}_{EGLO} (\epsilon_{\gamma}) = 8.674 \cdot 10^{-8}
\sigma_{r} \Gamma_{r} \times$$
\begin{equation}
\times \Bigg[ \frac{ \epsilon_{\gamma} \Gamma_{k} (\epsilon_{\gamma},T_{f})}
{(\epsilon_{\gamma}^2 - E_{r}^2)^2 + (\epsilon_{\gamma} \Gamma_{k}
(\epsilon_{\gamma}, T_{f}))^2 } +
 0.7  \Gamma_{k} (\epsilon_{\gamma} = \frac{0, T_{f})}{E_{r}^3 } \Bigg] ,
\label{gamma14}
\end{equation}
\noindent where $T_f$ is the temperature of the final state.
The  energy-dependent width $\Gamma_{k} (\epsilon_{\gamma},T_f)$ is taken proportionally
to the collisional damping width in Fermi-liquid scaled by an empirical \cite{KMF83}
 function
$K(\epsilon_{\gamma})$:
\begin{equation}
\begin{aligned}
 \Gamma_{k}(\epsilon_{\gamma},T)=K(\epsilon_{\gamma})
 \Gamma_{r}
 \left[\epsilon_{\gamma}^2+(2\pi T)^2 \right]/ E_{r}^2 ,\\
K(\epsilon_{\gamma}) = \kappa+(1-\kappa)
(\epsilon_{\gamma}- \epsilon_{0})/ (E_{r}-\epsilon_{0}).
\end{aligned}
\end{equation}
Dipole strength of the GFL model \cite{MD2000},
$\overleftarrow{f}_{E1} \equiv \overleftarrow{f}_{GFL}$, can be
given in the  following form for spherical nuclei
\begin{eqnarray}
\overleftarrow{f}_{GFL}(\epsilon_{\gamma}) &=& 8.674 \cdot 10^{-8}
\sigma_{r} \Gamma_{r} {K _{GFL}\epsilon_{\gamma}
\Gamma_{m}(\epsilon_{\gamma},T_{f}) \over ( \epsilon_{\gamma}^2 -
E_{r}^2)^2 + K _{GFL} (\Gamma_{m}(\epsilon_{\gamma},T_{f})
\epsilon_{\gamma})^2},
\label{gamma10} \\
K_{GFL} & =& \sqrt{E_{r}/E_{0}} =
(1+F^{\prime}_{1}/3)^{1/2}/(1+F^{\prime}_{0})^{1/2} = 0.63 ,
\nonumber
\end{eqnarray}
where $F^{\prime}_{0}$  and  $F^{\prime}_{1}$ are the Landau
parameters of the quasiparticle interaction in isovector channel
of the Fermi system; $F^{\prime}_{0}=1.49$  and
$F^{\prime}_{1}=-0.04$; $E_{0}$ is the average energy of one-particle
 one-hole states forming GDR. The
Eq.(\ref{gamma10}) is an extension of  original expression of the
GFL model \cite{MD2000} for wide range of the gamma-ray
energies: term $
K_{GFL}(\Gamma_{m}\epsilon_{\gamma})^2$ is
added to the denominator of the Eq.(\ref{gamma10}) to avoid
singularity of the GFL approach near GDR-energy in a way  similar
to  the  other models for E1 strength but with additional factor
$K_{GFL}$ in order to keep  standard relationship between
value of the RS function at the GDR energy and peak value
$\sigma_{r}$ of the photoabsorption cross-section.

The width $ \Gamma_{m}$ in (\ref{gamma10}) is taken  as a sum of a
collisional damping width, $\Gamma_{C}$, and a term,
$\Gamma_{dq}$, which simulates the fragmentation width,
\begin{equation}
\Gamma_{m}(\epsilon_{\gamma}, { T}) =
\Gamma_{C}(\epsilon_{\gamma}, { T}) +
\Gamma_{dq}(\epsilon_{\gamma}) , \ \ \ \
\Gamma_{C}\equiv C_{coll} \left(\epsilon_{\gamma}^{2}+ 4 \pi^{2}{T}^{2} \right)
, \label{gamma11}
\end{equation}
$C_{coll}=16m\cdot\sigma(np)/4\pi^2 9\hbar^2$.
 Here, a magnitude of the in-medium  cross section $\sigma (np)$
is taken   proportional to the value of the free space cross
section $\sigma_{f}(np) = 5~fm^2$ (near Fermi surface) with a factor
$F$. The form of collisional component corresponds to damping width
in infinite the Fermi-liquid and $\Gamma_{dq}$ is considered as
resulted from spreading GDR over surface
quadrupole vibrations due to dipole-quadrupole interaction:
\begin{equation}
\Gamma_{dq}(\epsilon_{\gamma}) = C_{dq} \epsilon_{\gamma} \mid
\bar{\beta}_{2}\mid \sqrt{1 + {E_{2} \over \epsilon_{\gamma}}} =
C_{dq} \sqrt{\epsilon_{\gamma}^{2}\bar{\beta}_{2}^{2} +
\epsilon_{\gamma}s_{2}}, \ \ \ s_{2}=E_{2}\bar{\beta}_{2}^{2} .
\label{gamma12}
\end{equation}
Here, $C_{dq}= (5\ln{2}/ \pi)^{1/2}= 1.05$; $E_{2}$ is energy of
the first excited vibrational $2^{+}$ state;
$\bar{\beta}_{2}$ is effective deformation parameter
of nuclear surface. It is determined by reduced electric photoabsorption
transition rate
for the ground state to $2^{+}$ state
transition \cite{Raman2001}.
The global parameterization \cite{Raman2001} is adopted
for $s_{2}$:
$s_{2} \equiv E_{2}\bar{\beta}_{2}^{2} = 217.16/A^{2}$ ,
when experimental data for even-even nuclei and in the case of odd and
odd-odd nuclei.

It should be noted that there
is not consistency of GFL model  with detailed balance principle
in systems with constant temperature and the collisional
components of the damping width of the GFL model can have negative
values in some deformed nuclei if magnitude $C_{coll}$ in
Eq.(\ref{gamma11}) is found from fitting width $\Gamma_m$
to experimental value of GDR width.

Expression for dipole RS within MLO model, $\overleftarrow{f}_{E1}
\equiv \overleftarrow{f}_{MLO}$, is obtained by calculating the
average radiative width in nuclei with microcanonically
distributed initial states (\cite{plu99a}-\cite{plu2001},\cite{plu90})).
In spherical nuclei the dipole radiative strength function is
proportional to the strength function for nuclear response on
dipole field  with frequency $\omega = \epsilon_{\gamma}/\hbar$.
An analytical semiclassical expression
for nuclear response function in cold and heated nuclei with
excitation of GDR  and theory-supported expressions for damping
are used.
As the result the radiative
strength function within MLO model has the following general form
(in MeV)
\begin{equation}
\label{MLO}
\overleftarrow{f}_{MLO} (\epsilon_{\gamma}, {T_f}) =
8,674 \cdot
10^{-8} \sigma_{r} \Gamma_{r}L
(\epsilon_{\gamma },T_f)
 \frac{\epsilon_{\gamma} \Gamma(\epsilon_{\gamma},{T_f})}
 { (\epsilon_{\gamma}^2 - E_{r}^2)^2 + (\Gamma(\epsilon_{\gamma},{T_f})
\epsilon_{\gamma})^2},
\end{equation}
where scaling factor $ L( \epsilon_{\gamma },T_f)=
1/[1-exp(-\epsilon_{\gamma}/T_f)]$ is essential for
low-energy radiation and leads to consistency of
the Eq.(\ref{MLO}) with detailed balance principle in systems
with constant temperature \cite{ER1993,plu90}.

Three variants of the modified Lorentzian model (MLO1, MLO2 and
MLO3) are considered below. They include different expressions
for damping width $\Gamma(\epsilon_{\gamma},{T_f})$. The
expression for MLO1 approach has the following
 form (\cite{plu2000a}-\cite{plu2001})
\begin{equation}
       \Gamma_{MLO1}(\epsilon_{\gamma},T_f) \simeq
       \beta  \gamma_{c}(\epsilon_{\gamma},T_f)
       \frac{E_{r}^{2}+E_{0}^{2}}{(E_{r}^{2}-E_{0}^{2})^2
       +(\gamma_{c}(\epsilon_{\gamma},T_f)\epsilon_{\gamma})^2},
      \end{equation}
       $$\beta = \left(1+\frac{E_{r}^{2}}{E_{0}^{2}}\right)^2 \frac{E_{0}^{2}}{2},\;\;\;\;\;
       \gamma_{c}(\epsilon_{\gamma},T) =
       \frac{2 \hbar }{ \tau_{c}(\epsilon_{\gamma},T)}.$$

The energy $E_{0}$ of particle-hole state is taken as equal the
harmonic oscillator energy $\hbar\omega_{0}={41}/A^{1/3},\ \ MeV$.
In this model the relaxation time within doorway state mechanism \cite{PGK2001}
 is used
\begin{equation}
 \label{eF}
 \frac{ \hbar}{\tau_{c}(\epsilon_{\gamma},T)} =
 b \left(\epsilon_{\gamma}+ U \right) ,\;\; \;
 b=\frac{E_{r}}{4\pi} \frac{F}{\alpha},\;\;\;
 \alpha = \frac{9\hbar / 16 m}{\sigma^{free}(np)},\;\;\;
 F=\frac{\sigma(np)}{\sigma^{free}(np)}.
 \end{equation}
The damping widths  in MLO2 and MLO3 models  are taken in approximation
of independent sources of dissipation \cite{KPS96} as a sum of the collisional damping width
$\Gamma_C$ and a term which simulate the fragmentation component of the
width:
$\Gamma(\epsilon_{\gamma},T_f)=\Gamma_C(\epsilon_{\gamma},T_f)+
\Gamma_F(\epsilon_{\gamma})$. The component $\Gamma_{C}$ is
inversely proportional to  the collision relaxation time $\tau$ in
isovector channel at dipole distortion of the Fermi surface \cite{PGK2001}.
The magnitude
of $\Gamma_{F}$  is taken proportionally to the wall formula value \cite{MSKEH77}
$\Gamma_{w}$ with a scaling factor $k_{s}$:
$ \Gamma_{F}(\epsilon_{\gamma}) =
k_{s}(\epsilon_{\gamma}) \Gamma_{w} ,
\Gamma_{w} = 36.43 \cdot A^{ - 1/3}~(MeV)$.
As a result, the
expression for the damping width has the following form,
\begin{equation}
\label{e59} \Gamma_{MLO2,3}(\epsilon_{\gamma},T_f) =
\hbar/\tau_{c}(\epsilon_{\gamma},T_f)+ k_{s}\Gamma_{w}.
\end{equation}

The energy-dependent power approximation is adopted for simplicity
for factor $k_{s}$ :
$k_{s}(\epsilon_{\gamma}) = k_{r} + (k_{0}-k_{r})
 \vert (\epsilon_{\gamma} - E_{r})/E_{r}\vert^{n_{s}}$, when
 $\epsilon_{\gamma} < 2 E_{r}$, and
$k_{s}(\epsilon_{\gamma})=k_{0}$ if $\epsilon_{\gamma} \geq 2 E_{r}$,
where the quantities $k_{0} \equiv
k_{s}(\epsilon_{\gamma}=0)$, $k_{r} \equiv
k_{s}(\epsilon_{\gamma}=E_{r})$ determine the contribution of the
"wall" component to the width at zero  energy and GDR-energy,
respectively. The value of the $k_{r}$ is obtained from fitting
the GDR width $\Gamma_{r}$ at zero temperature by the expression
(\ref{e59}) with $\epsilon_{\gamma} = E_{r}$. The
quantities $k_{0}$ and $n_{s}$  are  some parameters which are
obtained below from condition of correct description of a
general behaviour of the experimental gamma-decay strengths.
The collisional damping width $\Gamma_C$ is
taken in form given by Eq.(\ref{eF}) in the case of MLO2 model.
The expression for relaxation time  according to
the Fermi-liquid approach is used for MLO3 model:

      \begin{equation}
      \label{o15}
       \frac{\hbar}{\tau_{c}(\epsilon_{\gamma},T)}\equiv
       \frac{F}{\alpha}\left[ \left(\epsilon_{\gamma}/2\pi \right)^2 +T^2 \right].
      \end{equation}


\section{Analysis of the dipole strength function calculations}

 The RS functions were calculated  on following nuclei: $^{198}Au$,
$^{168}Er$, $^{156}Gd$, $^{158}Gd$ $^{146}Nd$, $^{144}Nd$,
$^{196}Pt$, $^{148}Sm$, $^{182}Ta$,$^{174}Yb$, $^{172}Yb$,
$^{90}Zr$, $^{139}Ba$, $^{137}Ba$, $^{120}Sn$, $^{106}Pd$.
The shape of the RSF within all models was investigated and compared with
experimental data; the $\chi^2$-criteria was used.
In Fig.1 the shape of the RSF
in $^{146}Nd$ is given as an example.
Here, only the MLO2 model is shown, because of similarity of the RSF
for different MLO approaches.
The following global set of parameters was used
$F=1,~ k_{s}=0.3,~ n_{s}=0.1.$

This set can be recommended for calculations within MLO
models. The optimal parameters for
each nuclear can be used  for more precise calculation.
Calculation within GFL and MLO models are rather similar,
and they are in close agreement with experimental data.

The GFL and MLO models can be also used to  estimate the M1
strength $\overleftarrow{f}_{M1}(E_{\gamma})$ in a wide interval
of $\gamma$-ray energies on the base of the ratio
$R=\overleftarrow{f}_{E1}(B_{n})/\overleftarrow{f}_{M1}(B_{n})$ for
the E1 and M1 strength functions at neutron binding energy $B_{n}$.
Then the M1 strength function is calculated by the
following relationship
\begin{equation}
\overleftarrow{f}_{M1}(E_{\gamma})={ \overleftarrow{f}_{E1}(B_{n})
\over R} {\phi_{M1}(E_{\gamma}) \over \phi_{M1}(B_{n})} ,
\label{gamma36}
\end{equation}
where $\phi_{M1}(E_{\gamma})$ is a function describing shape of
the dipole magnetic radiative strength function.
The magnitude of the $R$ can be obtained from experimental
data or systematics \cite{RIPL}:
$R= {\overleftarrow{f}_{E1}(B_{n}) /
\overleftarrow{f}_{M1}(B_{n})} = 0.0588 \cdot A^{0.878} , \ \ \
B_{n} \approx 7~MeV . \label{gamma37}$

Two models \cite{RIPL} for a function $\phi_{M1}$ are usually used:

1)~$\phi_{M1}(E_{\gamma})=const$  according to single- particle
model;

2)~$\phi_{M1}(E_{\gamma})$ is taken in form of the SLO
model(\ref{SLO}) corresponding to spin-flip giant resonance
mode with the following global values \cite{RIPL} for energy
and damping width (in $MeV$): $E_{r}=41 \cdot A^{-
1/3}$, $\Gamma_{r} = 4$.

The radiative strength functions for $E1+M1$ transitions
were calculated for
the following nuclei: $^{200}Hg$, $^{196}Pt$, $^{198}Au$,
$^{192}Ir$, $^{190}Os$, $^{183}W$, $^{182}Ta$, $^{181}Hf$,
$^{177,176}Lu$,  $^{174}Yb$,  $^{170}Tm$, $^{168}Er$, $^{166}Ho$,
$^{164}Dy$, $^{160}Tb$, $^{158}Gd$, $^{150}Sm$, $^{146}Nd$,
$^{140}La$, $^{139,138,137}Ba$, $^{128}I$, $^{125,124}Te$,
$^{114}Cd$, $^{90}Zr$, $^{80}Br$.
The experimental data were taken from Ref.\cite{VS}.

In Fig.2 the RS functions in $^{174}Yb$ are shown. Here, the
global parameter set is used for MLO2 model.
It can be seen that MLO and GFL models are also
usable in case of the $E1+M1$ emission.

\vspace{-0.5cm}

\section{Conclusions}

The numerical studies led to the following conclusions
on the description of the $E1$ and $E1+M1$ gamma-decay strength
functions by the simple analytical models.
The calculations by the MLO and GFL models are in more
close agreement with experimental data than that ones within the EGLO and SLO
methods at the energies $\epsilon_{\gamma}\lsim 4.5~MeV$. The
overall comparison between  calculations within the MLO, EGLO, SLO
and GFL models  and experimental data showed that the MLO and GFL
approaches provide a rather reliable method of the $\gamma$~-
decay strength description in a relatively wide energy
interval ranging from zero gamma-ray energy to values above GDR
peak energy. The MLO and GFL methods are not time consuming. They
can be applicable for calculations and  predictions of the
statistical contribution to the dipole strength functions as well
as for extraction of the GDR parameters of heated nuclei with
small errors with use of the $\gamma$-emission data.

This work is supported in part by the IAEA
(Vienna) under contract 302-F4-UKR-11567.


\newpage
\pagestyle{empty}

\begin{figure}
\begin{center}
\hspace{-10mm}
 \includegraphics[width=0.8\textwidth,clip]{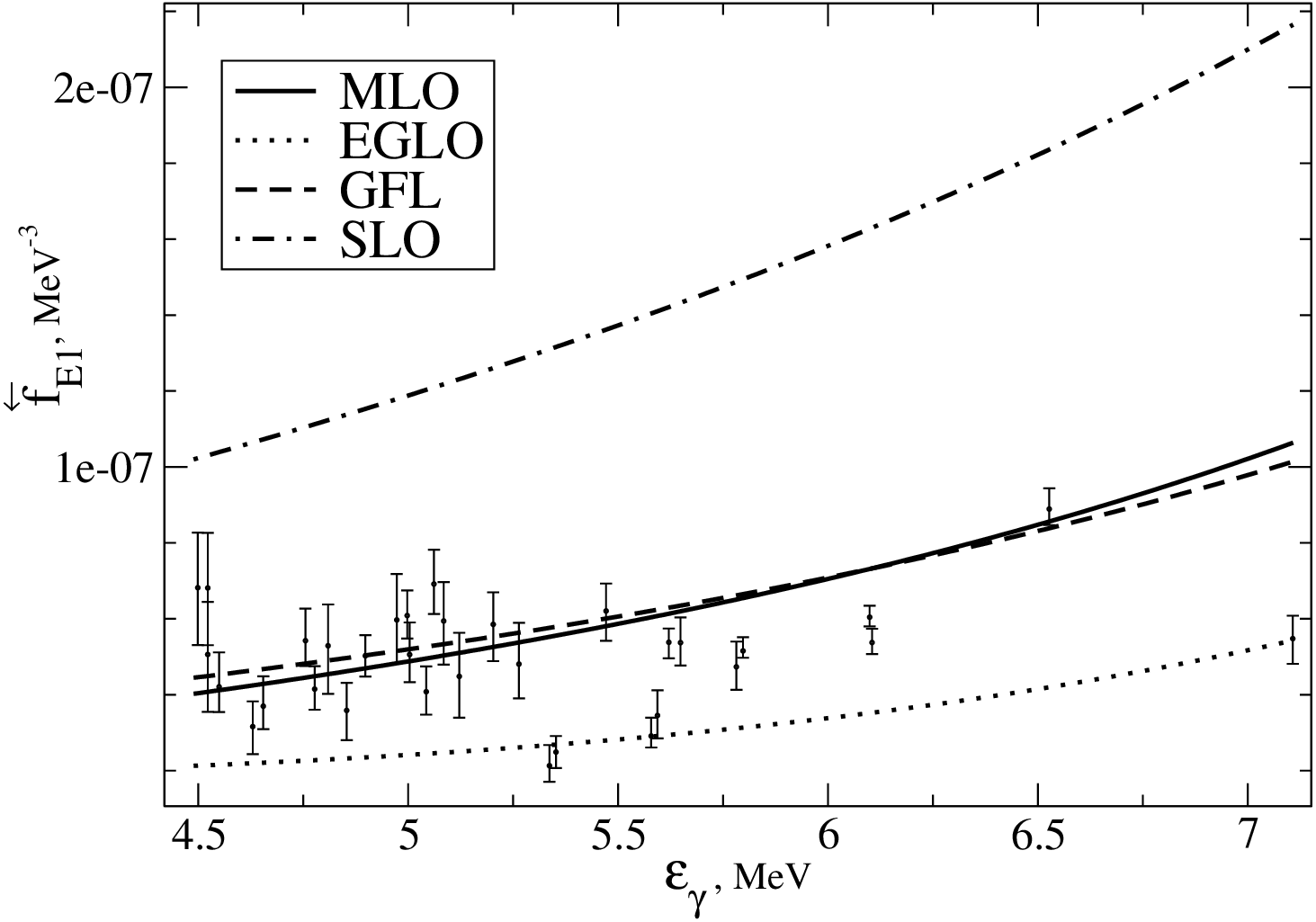}
\vspace{10mm}
 \parbox[t]{0.9\textwidth}{Fig.1: The E1 gamma-decay strength functions
 versus gamma-ray energy for $^{146}Nd$;  $F=1$, $k_{s}=0.3$, $n_{s}=0.5$.}
 \end{center}
\end{figure}

\begin{figure}
\begin{center}
\hspace{-10mm}
 \includegraphics[width=0.8\textwidth,clip]{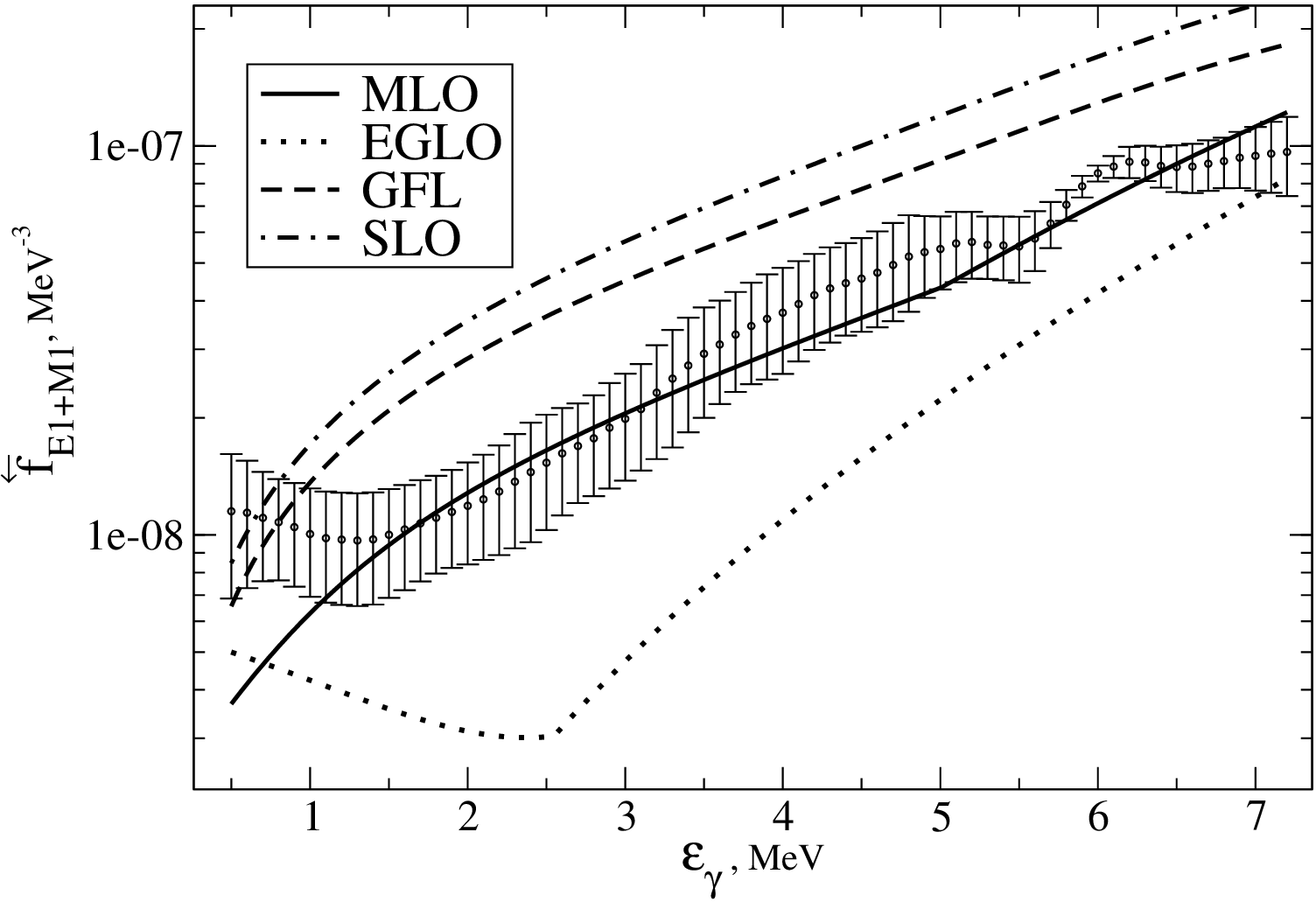}
\vspace{10mm}
 \parbox[t]{0.9\textwidth}{Fig.2: The $E1+M1$ gamma-decay
strength functions versus gamma-ray energy for  $^{174}Yb$.}
 \end{center}
\end{figure}

\end{document}